\documentstyle[11pt,a4]{article}  

\newcommand{\sect}[1]{\setcounter{equation}{0}\section{#1}}
\newcommand{\subsect}[1]{\subsection{#1}}

\newcommand{\vs}[1]{\rule[- #1 mm]{0mm}{#1 mm}}

\newcommand{\lbl}[1]{\label{eq:#1}}
\newcommand{\rf}[1]{(\ref{eq:#1})}
\newcommand{\nn}{\nonumber}
\newcommand{\Ch}{{C^{(2)}}} 
 
\newcommand{\be}{\vs{2}\begin{equation}}
\newcommand{\ee}{\vs{2}\end{equation}}

\newcommand{\bea}{\begin{eqnarray}}
\newcommand{\ena}{\end{eqnarray}}
\newcommand{\nnbea}{\begin{eqnarray*}}
\newcommand{\nnena}{\end{eqnarray*}}
\newcommand{\leqa}{\lefteqn}
 
\newcommand{\tld}{\widetilde}
\newcommand{\lra}{\ \longrightarrow\ }

\newcommand{\ovl}[1]{\overline{#1}}

\newcommand{\sm}[2]{\textstyle{\frac{#1}{#2}}\displaystyle}

\newcommand{\dps}{\displaystyle}

\newcommand{\bz}{{\ovl{z}}}
\newcommand{\bc}{{\ovl{c}}}

\newcommand{\bZ}{{\ovl{Z}}}
\newcommand{\yz}{y_{z}}
\newcommand{\ybz}{\ovl{y}_{\bz}}

\newcommand{\YZ}{Y_{Z}}

\newcommand{\YbZ}{\ovl{Y}_{\bZ}}

\newcommand{\zbz}{(z,\bz)}
 
\newcommand{\zbzt}{(z,\bz,\theta)} 
\newcommand{\zy}{(z,y)}

\newcommand{\zY}{(z,Y)}
\newcommand{\zYt}{(z,Y,\theta)} 
\newcommand{\ZY}{(Z,Y)}
\newcommand{\Zn}{Z^{(n)}}
\newcommand{\Zr}{Z^{(r)}}

\newcommand{\YbY}{(\YZ,\YbZ)}
\newcommand{\cR}{{\cal R }}
\newcommand{\cW}{{\cal W }}
 
\newcommand{\cU}{{\cal U }}
\newcommand{\sS}{{\cal S }}
\newcommand{\cK}{{\cal K }}
 
\newcommand{\cF}{{\cal F }}
\newcommand{\cC}{{\cal C }}

\newcommand{\cD}{{\cal D }}
\newcommand{\cL}{{\cal L }}

\newcommand{\cT}{{\cal T }}
\newcommand{\cM}{{\cal M }}
\newcommand{\cS}{{\cal S }}
\newcommand{\cV}{{\cal V }} 

\newcommand{\cX}{{\cal X }}
\newcommand{\cY}{{\cal Y }}

\newcommand{\cP}{{\cal P }} 

\newcommand{\prt}{\partial}

\newcommand{\lambdapi}{\lambda_z^{Z^{p_i}}}

\newcommand{\bprt}{\ovl{\partial}}

\newcommand{\ZBZ}{(Z\zbz,\ovl{Z}\zbz)}
\newcommand{\Zzero}{Z_0\zbz} 
\newcommand{\Zdzero}{Z^{(2)}_0\zbz}  
\newcommand{\ZBZzero}{(Z_0\zbz,\ovl{Z}_0\zbz)}  
 
\newcommand{\ZBZt}{(Z\zbzt,\ovl{Z}\zbzt)} 
\newcommand{\ZBZn}{(Z^{(n)}\zbz,\ovl{Z}^{(n)}\zbz)}
\newcommand{\ZBZr}{(Z^{(r)}\zbz,\ovl{Z}^{(r)}\zbz)} 
\newcommand{\ZBZdue}{(Z^{(2)}\zbz,\ovl{Z}^{(2)}\zbz)}
\newcommand{\ZBZduezero}{(Z^{(2)}_0\zbz,\ovl{Z}^{(2)}_0\zbz)} 
\newcommand{\ZBZduet}{(Z^{(2)}\zbzt,\ovl{Z}^{(2)}\zbzt)}

\newcommand{\wdelta}{\widetilde{\delta}}
 
\newcommand{\Zbn}{\bZ^{(n)}} 

\newtheorem{guess}{Theorem}[section]
\newtheorem{Remark}{Remark}[section]
\newtheorem{statement}{Statement}[section] 
\newtheorem{Ansatz}{Ansatz}[section]

\begin{document}

\pagestyle{empty}


\font\fifteen=cmbx10 at 15pt
\font\twelve=cmbx10 at 12pt

\begin{titlepage}

\begin{center}
\renewcommand{\thefootnote}{\fnsymbol{footnote}}

{\twelve Centre de Physique Th\'eorique\footnote{
Unit\'e Propre de Recherche 7061
}, CNRS Luminy, Case 907}

{\twelve F-13288 Marseille -- Cedex 9}

\vskip 1.5cm

{\fifteen  The geometry of $\cW_3$ algebra: a twofold way\\[2mm] 
for the rebirth of a symmetry } 

\vskip 1.5cm

\setcounter{footnote}{0}
\renewcommand{\thefootnote}{\arabic{footnote}}

{\bf G. BANDELLONI} $^a$\footnote{e-mail : {\tt beppe@genova.infn.it}}
\hskip .2mm  and \hskip .7mm {\bf S. LAZZARINI} $^b$\footnote{and also
Universit\'e de la M\'editerran\'ee, Aix-Marseille II. e-mail : {\tt
sel@cpt.univ-mrs.fr} }\\[6mm] 
$^a$ \textit{Dipartimento di Fisica
dell'Universit\`a di Genova,}\\
{\it Via Dodecaneso 33, I-16146 GENOVA, Italy}\\
 and \\
{\it Istituto Nazionale di Fisica Nucleare, INFN, Sezione di Genova}\\
{\it via Dodecaneso 33, I-16146 GENOVA, Italy}\\[4mm]
$^b$ {\it Centre de Physique Th\'eorique, CNRS Luminy, Case 907,}\\
{\it F-13288 MARSEILLE Cedex, France} 
\end{center}

\vskip 1.cm

\centerline{\bf Abstract}

The purpose of this note is to show that $\cW_3$
algebras originate from an unusual interplay between the breakings of the
reparametrization invariance under the diffemorphism action on the
cotangent bundle of a Riemann surface. It is recalled how a set of
smooth changes of local complex coordinates   on the base space are
collectively related to a background  within a symplectic
framework. The power of  the method allows to
calculate explicitly some primary fields whose OPEs  generate the
algebra as explicit functions in the coordinates:   this is achieved
only if well defined conditions are satisfied, and new symmetries
emerge from the construction.  Moreoverer, when primary flelds
are introduced outside of a coordinate description the  $\cW_3$
symmetry byproducts acquire a good geometrical definition with
respect to holomorphic changes of charts.

\vskip 1.5cm

\noindent 1998 PACS Classification: Conformal field theory algebraic
structures 11.25 Hf.

\indent


\indent

\noindent{CPT--2000/P.4078}, Internet : {\tt www.cpt.univ-mrs.fr}

\end{titlepage}

\renewcommand{\thefootnote}{\arabic{footnote}}
\setcounter{footnote}{0}
\pagestyle{plain}
\setcounter{page}{1}

\newpage

\sect{Introduction}

It is hard to decide if it is more relevant, from the physical point of view, 
either the birth or the decease of a symmetry.

The origin of a simmetry clarifies its deep purest form of
manifestation  and discloses its harmonies, while the symmetry
breaking mechanisms, even if they display the physical world  faults, shed
some new light on the dynamical realization of the symmetry. Indeed the
breaking consistency conditions \cite{BRSsymm}prevent a wild
discharge of the symmetry constraints. Moreover if we are in presence
of several symmetry violation mechanisms,  a mutual liking of them
could generate a fascinating guide-line leading to a new symmetry.
  
The purpose of this note is to show that one of the possible origins
of $\cW_3$  \cite{OSSN,hull,many,torino,Sorella,Grimm,Ader1,Ader2}
algebras comes from an unusual interplay between  diffeomorphism
symmetry breakings between the base  space (with local complex
coordinates $\zbz$) and a well defined chain of spaces (with local
complex coordinates $\ZBZn$) whose construction preserves the symplectic
diffeomorphism invariance.

Indeed the $\cW$ algebras saga \cite{Tjin} comes from many sources:
 they were originally discovered within an OPE construction by
 Zamolodchikov \cite {Zam1,Zam2} as a natural extension of the
 Virasoro algebra; later on they were derived by Drinfeld and Sokolov
 \cite{DrinSok} through a reduction procedure, which is in fact
 (classically) simply a Poisson reduction in infinitely many dimensions,
 taking as Poisson structure the Kirillov-Poisson structure naturally
 associated to any Lie algebra   \cite{GelDik,7}.
 
The same algebras were derived within the  Korteveg de Vries hierarchy
of equations  an approach which generalized Toda system formalism
\cite{Perem} giving fundamental insights in the field of the 2D matter
physics\cite{Capp}, integrable models \cite{Gieres},
topological 2D field theory \cite{She92}, 2D conformal field
theory\cite{Bou}, 2D quantum gravity\cite{Hull1} as well as matrix
models \cite{7}.

So, having a lot of possible sources, one may suppose
that a common general trend could link the widespread 
variety of these physical and mathematical fields of research.

The two dimensional space exclusive feature is that it admits
an infinite dimensional diffeomorphism  group of transformations
\cite{BPZ,BK} for which higher spin extensions of the group of
representations are allowed. 
  
This fact was already sensed within the $\cW$
field of interest
 from the very 
beginning:  indeed in the instance of $\cW$ chiral sector,   the
transformation laws of the BRS ghosts, $\cC^{(r)}\zbz$, was
found to be
\begin{eqnarray}
\sS \cC^{(r)}\zbz =\sum_s  s\,\cC^{(s)}\zbz\prt \cC^{(r-s+1)}\zbz +
(\mbox{stuff}) .
\lbl{scint}
\end{eqnarray}
This gives a slight indication that this symmetry is 
involved with coordinate transformation laws (the so-called $\cW$ 
diffeomorphisms) \cite{Hull1,geom,deBoerGoeree,24,25,26}.

This ansatz addresses both the questions to which the $\ZBZr$ spaces are
locally diffeomorphic to the $\zbz$  base space and what could be the general
construction law which originates them.
 
Already Witten \cite{Wi1} pointed out, with argument of algebraic
 topology, that the use of  Poisson  brackets induces a kind of
 symplectic geometry,  and gave the conjecture that $\cW$ algebras
 could be related to  symplectomorphisms.  In a recent paper
 \cite{BaLa0} we have taken inspiration from this idea in an attempt
 to provide  an alternative approach to these algebras.

In our scheme a sequence of local complex coordinates
 $\ZBZn$ will be found. The
 latter are built up in such a way to preserve a  symplectic form, and turn
 out to be the images under canonical diffeomorphisms of the $\zbz$
 base.  Under the diffeomorphism action on the cotangent bundle,
these coordinate transformations give rise to BRS variations
 similar to that of Eq \rf{scint}   and introduce an alternative attempt to a
 $\cW$ algebra realization.  For example, some questions will find  an
  easy answer: first of all a Lagrangian formalism   can be
 naturally embedded and the role of complex structure
 \cite{geom,Zucchini} (along the lines of \cite{BPZ})  can be
 explained \cite{BaLa00}.  However not all the $\cW$ algebras that can
 be found in the literature are naturally explained in our
 formalism, but the $\cW_3$ algebras in Refs \cite{OSSN} and
 \cite{hull} acquire a particular role in our approach: on the one
 hand, it does not maintain  the reparametrization invariance
 previously introduced, 
 but on the other hand, the breaking mechanisms are fully under
 control in our treatment 
by imposing a  reparametrization invariance in a $\zbzt$ space, where
 $\theta$ is a constant  Grassmann variable which allows to manage in
 an algebraic way  the symmetry breakings with respect to the $\zbz$ background.
 
This problem is the subject of the paper, and it will be shown that 
in the most general case \cite{OSSN}, there
are two distinct mechanisms for the symmetry breakings,  whose accordance
produces a liking mechanism which generates a symmetry criterion.

On the other side, in the particular case of the $\cW_3$ of
\cite{hull} a surviving reparametrization invariance 
maintains its validity.    


As a byproduct of our approach we shall try to use the symmetry as a
firepower to  construct primary fields  (whose OPE gave historically
the origin to $\cW$ algebras) as {\it explicit} functions of the
$\ZBZn$ coordinates.  This construction will be possible, and in that
case an {\it explicit} reparametrization invariance will survive under
the diffeomorphism action as in the situation of \cite{hull}.

Otherwise \cite{OSSN} primary fields have to be introduced as
independent fields depending on the background $\zbz$ space.  This
point of view enhances for a good  geometrical definition of the
theory, in particular a well defined anomaly under holomorphic
changes  of charts will be found: this feature in not naively
shown in the first approach.

Our paper is organized as follows:

In Section 2  our approach to $\cW$ algebras from symplectomorphisms
is reviewed and the $\cW_3$ algebras are derived from a general construction.

In Section 3 the $\cW_3$ algebra is studied as a broking of the symmetry 
under reparametrizations on the base space in a well defined
 geometric scenario. In particular, 
we discuss the different aspects which emerge if  an explicit 
coordinate description of the algebra is required.

\sect{ $\cW$ algebras from symplectomorphisms: the chiral
$\cW_n$-gravity}

In this Section we sum up our approach to $\cW$ algebras 
in term of symplectomorphisms we have performed in Refs
\cite{BaLa0,BaLa00} to which we refer for more details .

The canonical 1-form $\Theta$ on the cotangent bundle $T^{*}$ writes in a  
local chart frame $\cU_{\zy}$

\begin{eqnarray}
{\Theta\arrowvert}_{\cU_\zy}=\biggl[\yz dz  +\ybz d\bz  \biggr]
\lbl{thetaz}
\end{eqnarray} 
In a frame $\cU_{\ZY}$, $\Theta$  will take the 
form 

\begin{eqnarray}
{\Theta\arrowvert}_{\cU_\ZY}=\biggl[\YZ dZ   + \YbZ    d\bZ  \biggr]
\lbl{thetay}
\end{eqnarray} 
 
The $(2,0)-(0,2)$ form 
$\Omega\equiv d\Theta$
 is globally defined both in $\cU_{\zy}$  and in $\cU_{\ZY}$  
\begin{eqnarray}
{\Omega\arrowvert}_{\cU_\zy} 
\equiv \biggl[d \yz \wedge dz  +  d \ybz \wedge d{\bz}\biggr
]
\lbl{omegaz}
\end{eqnarray}

\begin{eqnarray}
{\Omega\arrowvert}_{\cU_\ZY}=\biggl[d\YZ \wedge dZ  +
d\YbZ \wedge  d\bZ \biggr]
\lbl{omegaZ}
\end{eqnarray}

Recall that a change of charts is canonical if  in
$\cU_{\zy} \cap \cU_{\ZY}$ 
\begin{eqnarray}
{\Omega\arrowvert}_{\cU_\zy}={\Omega\arrowvert}_{\cU_\ZY}
\lbl{canonical}
\end{eqnarray}
which implies
\begin{eqnarray}
{\Theta\arrowvert}_{\cU_\zy}-{\Theta\arrowvert}_{\cU_\ZY}= d F
\end{eqnarray}
$F$ is  a function on  $\cU_\zy\cap \cU_\ZY$.
In the  $\zY$ plane,  an extra coordinate $\theta$ is now introduced 
which will turn out to be fundamental in the sequel. 
A generating function $\Phi\zYt$ is thus defined as:

\begin{eqnarray}
 d\Phi\zYt\equiv d \biggl (F\zYt + (\YZ Z\zYt+\YbZ \bZ\zYt \biggr)\nn\\
 =\biggl(\yz\zYt  dz + \ybz\zYt  d\bz\biggr)+
\biggl(d\YZ Z\zYt + d\YbZ  \bZ\zYt \biggr)
\lbl{Phi}
\end{eqnarray}
showing  that the mappings:
\begin{eqnarray}
\yz\zYt=\prt\Phi\zYt,\qquad
 \lbl{Phi1} Z\zYt=\frac{\prt}{\prt \YZ} \Phi\zYt
\end{eqnarray}
are canonical.

So that in the $\zYt$ chart, ${\Omega\arrowvert}_{\cU_{\zYt}}$
takes the elementary form:

\begin{eqnarray}
{\Omega\arrowvert}_{\cU_{\zYt}}
&=& d\YZ\wedge d_z Z\zYt +d\YbZ\wedge d_z \bZ\zYt\nn\\
&=&d_{\YZ}
 {\yz}\zYt \wedge dz +d_{\YbZ}{\ybz}\zYt \wedge d\bz\nn \\
&=& d_z d_Y\Phi\zYt
\lbl{omegafull}
\end{eqnarray}
where $d_Y $ and $d_z$ are the differentials operating in the $\YbY$ and $\zbz$
respectively. 

At this point a crucial remark is in order : in Eq \rf{omegafull}
the terms $d_{\YZ} Z \wedge d{\YZ}  +d_{\YZ}\bZ\wedge d{\YbZ} $ 
and $ d_{z}\yz \wedge dz + d_z \ybz \wedge d\bz $ will identically
vanish in 
$\Omega$.
So an infinitesimal variation of $Z\zYt$ in $\YZ$ does not modify, for fixed
$\zbzt$, the 2-form $\Omega$.

From this comment originates the important  Theorem \cite {BaLa0}.  
\begin{guess}
On the smooth trivial bundle $\Sigma\times\mbox{\bf
R}^2$,
the vertical holomorphic change of local coordinates, 
\begin{eqnarray}
Z(\zbzt ,\YbY) \lra Z(\zbzt, \cF(Y_Z),\YbZ),
\end{eqnarray}
where $\cF$ is a holomorphic function in $\YZ$, and the horizontal
holomorphic change of local coordinates,
\begin{eqnarray} 
\yz(z,\bz,\theta\YbY)\lra \yz(f(z),\bz,\theta\YbY),
\end{eqnarray}
where $f$ is a holomorphic function in $z$,
are both canonical transformations.
\end{guess}

So the local changes of complex coordinates $z \lra Z(\zbz,\YZ)$ and
$z \lra Z(\zbz,\YZ +d\YZ)$,
will be related to the same two form $\Omega$.
 
Expanding around, say, $\YZ=0,\YbZ=0$  the 
generating function $\Phi$  will be written as a formal power series in $Y$ :

\begin{eqnarray}
\Phi\zYt 
&=&
\sum_{n=1}^{n_{max}}\biggl[\YZ^n\, \Zn\zbzt\biggr]
+ \sum_{n=1}^{n_{max}} \biggl
[\YbZ^n\, \Zbn\zbzt\biggr]\nn\\
\lbl{phi}
\end{eqnarray}

The  extra coordinate $\theta$ previously introduced is now specified
as a constant Grassmann variable with Faddeev-Popov charge equal to
$-1$. One has the splitting

\begin{eqnarray}
\Phi\zYt=\Phi_0\zY+\theta\Phi_\theta\zY
\lbl{split}
\end{eqnarray}

where the $\Zr\zbzt,1\leq r\leq n$ are local coordinates defined by

\begin{eqnarray}
Z^{(r)}\zbzt&=&\frac{1}{r!}{\biggl(\frac{\prt}{\prt
\YZ}\biggr)}^r\Phi_\zYt\Arrowvert_{\YZ=0}=Z^{(r)}_0\zbz +\theta
Z^{(r)}_\theta\zbz 
\lbl{zr}
\end{eqnarray}
We introduce:  
\begin{eqnarray}
\lambda^{(r)}\zbzt &\equiv&\prt Z^{(r)}\zbzt 
=\prt\frac{1}{r!}{\biggl[\frac{\prt}{\prt \YZ}\biggr]}^r\Phi\zYt\nn\\
\lambda^{(r)}\zbzt\mu(r,\zbzt)&\equiv&\bprt Z^{(r)}\zbzt 
=\bprt\frac{1}{r!}{\biggl[\frac{\prt}{\prt \YZ}\biggr]}^r\Phi\zYt
\end{eqnarray}
 and we shall denote $\lambda^{(1)}\zYt =: \lambda\zYt$,
$\mu(1,\zYt)=:\mu\zYt$
 
Note that for a given level $r$ the  coordinate $\Zr\zbzt$ is related
to other ones with index less than $r$  by \cite{BaLa0} 

\begin{eqnarray}
\Zr\zbzt&=&\biggr[
\frac{1}{r!} {\biggl(\frac{\prt}{\prt \YZ}\biggr)}^{r}
(\Phi\zYt)\biggr]\arrowvert_{\YZ=0}\nn\\[2mm]
&=&\sum_{j=1}^r j! \prod_{i=1}^{m_j} \biggl[
\frac{{(\prt Z^{(p_i)}\zbzt)}^{a_i}}{{a_i}!}
\biggr]_{\bigg|_
{\tiny{\left\{
\begin{array}{c}
{\sum_i a_i =j},\
{\sum_i a_i p_i=r}\\[2mm]
{p_1>p_2>,\cdots,>p_{m_j}} 
\end{array}\right\}}} }
\cM^{(j)}\zbzt\nn \\
\lbl{Zdecomp}
\end{eqnarray}
for all $ 0\leq r\leq n$ and where the functions $\cM^j\zbzt$ are given by:
\begin{eqnarray} 
\cM^{(j)}\zbzt\equiv\frac{1}{j!}{\biggl(
\frac{1}{\lambda\zYt}\frac{\prt}{\prt\YZ}\biggr)}^j\Phi\zYt 
\arrowvert_{\YZ,\bar{\YZ}=0}\equiv \cM^{(j)}_0\zbz +\theta
\cM^{(j)}_\theta\zbz             
\lbl{cM}                              
\end{eqnarray}                             

The differential operator $\frac{1}{\lambda\zYt}\frac{\prt}{\prt\YZ}$ can be
formally defined using the usual Grassmaniann inverse procedure  

\begin{eqnarray}
\frac{1}{\lambda\zYt}
  \frac{\prt}{\prt\YZ}
  =
  \biggl(\frac{1}{\lambda_0\zY}-\theta\frac{\lambda_\theta\zY}
{{(\lambda_0\zY)}^2}\biggr)\frac{\prt}{\prt\YZ}  
  \equiv\cD_0\zY+\theta\cD_\theta\zY
  \end{eqnarray} 
so that for $1\leq r\leq n $
  \begin{eqnarray}
 \cM^{(r)}_0\zbz&\equiv&\frac{1}{r!}{\biggl[\cD_0\zY\biggr]}^r\Phi_0\zY\Arrowvert_{\YZ=0}
\nn\\
  \cM^{(r)}_\theta\zbz&\equiv&\frac{1}{r!}\biggl(
{\biggl[\cD_0\zY\biggr]}^r\Phi_\theta\zY\nn\\
  & +&
  \sum_{j=0\cdots r-1}\cD_0^{(n-j)}\zY\cD_\theta\zY\cD_0^j\zY 
  \Phi_0\zY\biggr)\Arrowvert_{\YZ=0}   
\end{eqnarray}

\bigskip

The previous geometrical structure allows us to derive now a
$\cW$-symmetry in terms of the algebra of diffeomorphisms  combining both the
diffeomorphism action and the canonical transformations via the BRS
machinery. Calling $\sS$ the BRS differential operator for the
diffeormorphism action on the cotangent bundle \cite{BRS,BaLa0}, according
to the decomposition \rf{split}, one has
   \begin{eqnarray}
\sS\Phi\zYt=\Lambda\zbzt
\equiv\Lambda_0\zY+\theta\Lambda_\theta\zY
\end{eqnarray}
where $\theta$ is a constant field with Faddeev-Popov charge equals to
$-1$, and we shall fix its BRS variation as 
$\sS \theta =-1$. One identifies
\begin{eqnarray}
\Lambda_0\zY&=& \sS\Phi_0\zY -\Phi_\theta\zY\nn\\
\Lambda_\theta\zY&=&\sS\Phi_\theta\zY 
\end{eqnarray}
and the nilpotency condition on the generating function,
$\cS^2 \Phi\zY=0$ means:
\begin{eqnarray}
\cS\Lambda_0\zY&=&\Lambda_1\zY\nn\\
\cS\Lambda_\theta\zY&=&0.
\lbl{slambda}
\end{eqnarray} 
So  for each $n$ we define the diffeomorphism action on
the local complex coordinate $Z^{(r)}$ by,

  \begin{eqnarray}
  \sS Z^{(r)}\zbzt \equiv \Upsilon^{(r)}\zbzt
  =\frac{1}{r!}
  {\biggl[\frac{\prt}{\prt\YZ}\biggr]}^r\biggl[\Lambda\zYt\biggr]
_{\bigg|_{\YZ=0}}, \qquad
  1\leq r\leq n 
  \lbl{szr} 
  \end{eqnarray}
which, once decomposed according to its $\theta$ content, gives:
\begin{eqnarray}
\sS Z^{(r)}_0\zbz- Z^{(r)}_\theta\zbz=\frac{1}{r!}{\biggl(\frac{\prt}{\prt
\YZ}\biggr)}^r\Lambda_0\zY\Arrowvert_{\YZ=0}\nn\\ 
\sS Z^{(r)}_\theta\zbz=-\frac{1}{r!}{\biggl(\frac{\prt}{\prt
\YZ}\biggr)}^r\Lambda_\theta\zY\Arrowvert_{\YZ=0} \ .
\lbl{szr0}
\end{eqnarray}
One can easily verify, that $\sS^2 Z^{(r)}\zbzt=0$   in the $\zbzt$
  space (as in Eq\rf{szr}), but note that $\sS^2 Z_0\zbz\neq 0$ in the
  $\zbz$ space 
  (as in Eq \rf{szr0}) which leads to a  diffeomorphism symmetry
  breaking through  the
  local smooth changes of complex coordinates $\zbz\lra \ZBZr$, while
  the one for $\zbz\lra 
  (Z^{(r)}\zbzt, \ovl{Z}^{(r)}\zbzt)$ holds its full validity.

This shows the important role played by the $\theta$ field,
 and how the $\theta$ decomposition can be managed in the present treatment: 
 in order to fix all the algebraic relations, calculations must be performed in
 the $\zbzt$  space, and only after the $\theta$ decomposition is performed. 
 To sum up the construction:

\begin{statement}

The introduction of the $\theta$ coordinate in the $\zbz$ space gives
rise to a superselection   sector where the BRS breaking terms of the
diffeomorphism transformations $\zbz \lra \ZBZr$ lie.

Moreover the consistency conditions of these breakings and the B.R.S
 behaviour of the $\theta$ field preserves the diffeomorphism symmetry  
 $\zbz\lra (Z^{(r)}\zbzt \ovl{Z}^{(r)}\zbzt)$. 

\lbl{statement1} 
\end{statement} 
\bigskip

The diffeomorphism ghosts in
the $\zbzt$ space, are introduced as usual \cite{BaLa0}
and can also be written with
respect to the $\theta$ decomposition,
\begin{eqnarray}
\cK^{(r)}\zbzt=\frac{\Upsilon^{(r)}\zbzt}{\prt Z^{(r)}\zbzt}\equiv
\cK^{(r)}_0\zbz+\theta\cK^{(r)}_\theta\zbz
\lbl{cKr}
\end{eqnarray}
with the following transformation laws
\begin{eqnarray}
\sS \cK^{(r)}\zbzt= \cK^{(r)}\zbzt\prt \cK^{(r)}\zbzt
\lbl{skr}
\end{eqnarray} 
and 
\begin{eqnarray}
\sS \cK^{(r)}_0\zbz&=& \cK^{(r)}_0\zbz\prt \cK^{(r)}_0\zbz
+\cK^{(r)}_\theta\zbz\\
\sS \cK^{(r)}_\theta &=&\cK^{(r)}_0\zbz\prt \cK^{(r)}_\theta\zbz
-\cK^{(r)}_\theta\zbz\prt \cK^{(r)}_0\zbz
\lbl{skr0}
\end{eqnarray}
respectively.

The nesting of the $\ZBZr$ spaces brings to a decomposition
\cite{BaLa0}  of the ghost $\Upsilon^{(r)}\zbzt$ into the spaces of
order  lower than $r$,

  \begin{eqnarray}
\Upsilon^{(r)}\zbzt 
&=&\sum_{j=1}^r j! \prod_{i=1}{m_j}\biggl[
\frac{{(\lambdapi\zbzt)}^{a_i}}{{a_i}!}
\biggr]_{\bigg|_
{\tiny{\left\{
\begin{array}{c}
{\sum_i a_i =j},\
{\sum_i a_i p_i=r}\\[2mm]
{p_1>p_2>\cdots>p_{m_j}} 
\end{array}\right\}}} }
 \cY^{(r)}\zbzt \quad 
\lbl{kn}
\end{eqnarray}

where it has been defined for $1\leq r\leq n$
 \begin{eqnarray} 
 \cY^{(r)}\zbzt
 &=&\biggl(\cC^{(r)}\zbz+\theta \cX^{(r)}\zbz\biggr)\nn\\ 
\cC^{(r)}\zbz
&=&\frac{1}{r!}{\biggl[\cD_0\zY\biggr]}^r\Lambda_0\zY\Arrowvert_{\YZ=0} \nn\\
  \cX^{(r)}\zbz &=&
  \biggl( \frac{1}{r!}{\biggl[\cD_0\zY\biggr]}^r\Lambda_1\zY +
  \Sigma^r_\theta\zY\Lambda_0\zY\biggr)\Arrowvert_{\YZ=0}   \nn\\
\lbl{chi}
  \end{eqnarray}

With some tedious but straitforward algebraic manipulations
 the B.R.S variations of the previous fields can be computed by
using:
\begin{eqnarray}
\biggl[ \sS, \cD\zbz \biggr]=-\biggl( \cC\zbz \prt \ln \lambda_0\zbz +\prt
\cC\zbz +\frac{\lambda_\theta\zbz}{\lambda_0\zbz} \biggr) \cD\zbz
\lbl{SD}
\end{eqnarray}
(from now on the summation procedure is explicitly expressed:
repeated indices do not mean summation). One obtains as said
in the introduction variations of the type of Eq\rf{scint},
     
\begin{eqnarray}
\sS\cC^{(r)}\zbz& =&\sum_{s=1}^r s\cC^{(s)}\zbz \prt \cC^{(r-s+1)}\zbz
+\cX^{(r)}\zbz, \qquad r<n
 \lbl{scr} 
\end{eqnarray}
and, to insure nilpotency
\begin{eqnarray}
\sS \cX^{(r)}\zbz&=& \sum_{s=1}^r \biggl (
 s\cC^{(s)}\zbz \prt\cX^{(n-r+1)}\zbz\biggr)
- s\cX^{(s)}\zbz \prt \cC^{(r-s+1)}\zbz\nn\\[-5mm] 
&{}&\lbl{scr1}\\ 
 &=&- \sum_{s=1}^r \biggl((r-s+1)\cX^{(r-s+1)}\zbz \prt \cC^{(s)}\zbz
- s\cC^{(s)}\zbz \prt\cX^{(r-s+1)}\zbz\biggr)\nn
\end{eqnarray}

The $\cX^{(r)}\zbz$ fields are still undetermined objects: our aim
is to fix a recipe in order to identify them as functions of the $\cC^{(r)}\zbz$
ghost fields; we now show that if we impose by hand 
\begin{eqnarray}
\cX^{(n)}\zbz =0
\lbl{chin}
\end{eqnarray}
 then all the $\cX^{(r)}\zbz,\ r<n$ fulfill our requirements. Indeed
 from the fact that the ghost of maximum order $\cC^{(n)}\zbz$ has no
 breaking term, that is:  
 \begin{eqnarray}
 \sS\cC^{(n)}\zbz =\sum_{s=1}^{s=n} s\cC^{(s)}\zbz \prt \cC^{(n-s+1)}\zbz
 \lbl{scn} 
 \end{eqnarray}
so we can derive:
 \begin{eqnarray}
   \sS^2\cC^{(n)}\zbz =
-\theta\biggl[ 
 \sum_{s=1}^{n-1} s\cX^{(s)}\zbz \prt \cC^{(n-s+1)}\zbz
 -\sum_{s=2}^n s\cC^{(s)}\zbz \prt \cX^{(n-s+1)}\zbz\biggr]=0 
 \lbl{s2cn} 
 \end{eqnarray}
and try the more general solution as a differentail polynomial in the
ghost fields
\begin{eqnarray*}
\cX^{(n-r+1)}\zbz=\sum_{r=2}^n\sum_{r',t,m,m',l\geq 0}
\bprt^{r'}\prt^r\cC^{(t)}\zbz\bprt^{m'}\prt^m
\cC^{(l)}\zbz\cT^{(n-s+1-l-t+m+r),(r'+m')}\zbz
\lbl{X}
\end{eqnarray*} 
we obtain (there is no summation):
\begin{eqnarray}
\leqa{\cX^{(n-r+1)}\zbz =
\cC^{(r)}\zbz\prt\cC^{(r)}\zbz\cT^{(n-3r+2)}\zbz}  \nn\\
&& +\; \alpha_{(n-r+1)}\biggl(\prt\cC^{(r)}\zbz
\prt^2\cC^{(r)}\zbz-\frac{r}{n+1}\cC^{(r)}\prt^3\cC^{(r)}\zbz 
\biggr)
\delta_{(n-r+1)}^{(2r+3)}
\end{eqnarray}
 So the existence of $\alpha$ terms is possible only if 
 the condition $n-3r +4=0$ is solved by $n$ and $r$ integers.  
This is achieved for example for $n=r=2$ which turns out to be the
 $\cW(3)$ case. Rewriting the above solution for $1\leq r<n$ as

\begin{eqnarray}
\leqa{\cX^{(r)}\zbz =
\cC^{(n-r+1)}\zbz\prt\cC^{(n-r+1)}\zbz\cT_{(2n-3r+1)}\zbz } \\
&&+\alpha_{(r)}\biggl(\prt\cC^{(n-r+1)}\zbz
 \prt^2\cC^{(n-r+1)}\zbz - \frac{n-r+1}
{n+1}\cC^{(n-r+1)}\prt^3\cC^{(n-r+1)}\zbz  
\biggr) \delta^{2(n-r+1)-3}_{(r)} \nn
\end{eqnarray}
we then substitute into Eq\rf{scr} in order to get the properties of
 coefficients $\cT^{}$.

The terms ot the type $\cC^{(\cdots)}\prt\cC^{(\cdots ')}$(Stuff)
fix the B.R.S variations of $\cT_{(2n-3r+1)}\zbz$
the other terms must cancel. 
Let us first consider the monomials $\cC^{(\cdots)}\cC^{(\cdots
')}\prt^2\cC^{(\cdots ")}$. The terms: 
 
  \begin{eqnarray} 
& 1)&
   \sum_{s=1}^{r}\biggl[s\cC^{(s)}\zbz 
\cC^{(n-(r-s+1)+1)}\zbz\prt^2\cC^{(n-(r-s+1)+1)}
\zbz\cT_{(2n-3(r-s+1)+1)}\zbz \biggr]\nn\\ 
\nn\\ 
&2)&
\cC^{(n-r+1)}\sum_{s=1}^{n-r+1}
\biggl[ s\cC^{(s)}\prt^2\cC^{(n-r-s+2)}\biggr]\cT_{(2n-3r+1)}
\end{eqnarray} 
cancel  each  other  only for   $s=1$
 while for  $s>1$ they give inconstencies; so a priori we have to put
 in the term 1) $r=1$.  In this case this term collapses into
\begin{equation}
\cC^{(1)}\zbz\cC^{(n)}\zbz\prt^2\cC^{(n)}\zbz\cT_{(2n-2)}\zbz
\lbl{r=1}
\end{equation}
which is the first term(s=1)  of the series 2).
The remaining terms in the latter can be dropped out 
giving prescription for the parameter $n$.

The ``a priori logical choice'' would be  $n=1$ which reproduces
 the previous term, and trivializes the $\cW$ content. But
it has to be noted that if $n=2$ the top term of \rf{r=1}  is
zero for obvious Faddeev-Popov reason and the cancellation mechanism is
then achieved. 

 
For the highest value of $n$ the cancellation mechanism does not hold unless 
all the ghosts $\cC^{(j)}\zbz; j=2\cdots n-1$ are put to be zero and
we next repeat the  
same Faddeev-Popov trick for the $n$-th term of the sum in order to get 

\begin{eqnarray}
\sS \cC\zbz &=& \cC\zbz\prt\cC\zbz+
\cC^{(n)}\zbz\prt\cC^{(n)}\zbz\cT_{(2n-2)}\zbz \nn\\
&&+\ \alpha\biggl(\prt\cC^{(n)}\zbz
 \prt^2\cC^{(n)}\zbz-\frac{n} {n+1}\cC^{(n)}\prt^3\cC^{(n)}\zbz 
\biggr)
\delta^{(n)}_{(2)}\nn\\[2mm]
\lbl{ct}
\sS\cC^{(n)}\zbz &=& \cC\zbz\prt\cC^{(n)}\zbz 
+n\cC^{(n)}\zbz\prt\cC\zbz
\lbl{grimm}
\end{eqnarray}
 
for $n\neq 2$ this algebra was found in \cite{grimm1}.
So we conclude:

\begin{statement} 

The most general algebra of the type  Eqs\rf{scr}\rf{scr1} with constraints 
Eq\rf{scn}\rf{chin} is given in the  Equations \rf{ct}\rf{grimm}.
In particular the $\cW_3$ algebra is the only one which contain an $\alpha$
dependent term. 

\lbl{Statement00} 
\end{statement}

\subsect{The chiral $\cW_3$-gravity algebra}

We specialize here to the case $n=2$ ($\cW_3$) whose peculiarity 
  among  all the other situations is stated just above :
it is the only case which admits  $\alpha\neq 0$.
We shall see that the presence (or the absence) 
 of a term proportional to $\alpha$ is fundamental for the geometrical setting
of the problem.  
 
For different values of $n$ the general discussion can be easily performed
along the lines   here for $n=2$.

>From \rf{scn}, let us recall that  the highest order $\cX$ term has
been put to zero in order to fix
all the lowest order terms; here for $n=2$ we have to impose: 

\begin{eqnarray}
\cX^{(2)}\zbz\equiv \frac{1}{2}\biggr[
\cD_0^2
\Lambda_1\zY-\frac{\lambda_\theta\zY}{\lambda_0\zY}\cD_0^2\Lambda_0\zY
-\cD_0\biggl(\frac{\lambda_\theta\zY}{\lambda_0\zY}\cD_0\Lambda\zY\biggr)
\biggr]\Arrowvert_{\YZ=0} = 0 \quad
\lbl{cX2}
\end{eqnarray}
This constraint has a twofold face. The first one concerns the implications
on the $\cC$'s algebra, as  in the previous Section. The second
involves the role that $\cX^{(2)}\zbz$ is supposed to play in the
symplectic formalism by exploiting all the geometrical aspects 
emerging from the vanishing of the previous formula.
Only the former will be discussed here, the latter being postponed to
the next Section.  

First of all, the condition $\cX^{(2)}\zbz=0$ implies:
\begin{eqnarray}  
\sS \cC^{(2)}\zbz
= \cC\zbz\prt\cC^{(2)}\zbz +2\cC^{(2)}\zbz\prt\cC\zbz
\lbl{sc2}
\end{eqnarray}
and setting $\cX:=\cX^{(1)}$, Eq\rf{s2cn} yields 

\begin{eqnarray}
\cX \zbz\prt\cC^{(2)}\zbz= 2\cC^{(2)}\zbz\prt \cX\zbz
\lbl{condX}
\end{eqnarray}
 which is solved by
\begin{eqnarray}
\leqa{\cX\zbz =-\cC^{(2)}\zbz\prt \cC^{(2)}\zbz\frac{16 }{3}\cT\zbz}\nn\\
&&+\alpha\biggl(
\prt\cC^{(2)}\zbz\prt^2\cC^{(2)}\zbz-
\frac{2}{3}\cC^{(2)}\zbz\prt^3\cC^{(2)}\zbz\biggr)
\lbl{X0}
\end{eqnarray}
where we have been forced to introduce a spin $(2,0)$-conformal field
$\cT\zbz$. Summing up, we find the algebra:
\begin{eqnarray}
\sS \cC\zbz&= &\cC\zbz\prt\cC\zbz
-
 \frac{16 }{3}\cT\zbz \cC^{(2)}\zbz\prt \cC^{(2)}\zbz   \nn\\
&+& \alpha\biggl( 
\prt\cC^{(2)}\zbz\prt^2\cC^{(2)}\zbz 
- \frac{2}{3}\cC^{(2)}\zbz\prt^3\cC^{(2)}\zbz\biggr)\\
\sS \cC^{(2)}\zbz&=&\cC\zbz\prt\cC^{(2)}\zbz+2\cC^{(2)}\zbz\prt\cC\zbz \nn
\lbl{sc1}
\end{eqnarray}

\bigskip

Note that the $\cX\zbz$, as in our primary purpose in 
 Eq \rf{chi}, expresses the breaking  at     the level
 of the diffeomorphism algebra.
 It depends only on the $\cC^{(2)}\zbz$ ghost and its derivative; this
 moves us to make the following Remark: 

\begin{Remark}
In the limit of 
 $\cC^{(2)}\zbz $
  going
   to zero,
 we find the diffeomorphism
  algebra describing the ordinary $\zbz\lra \ZBZ$ reparametrization.\\
   So the ghost $\cC^{(2)}\zbz $ parametrizes the breaking of this symmetry at
the level of BRS  algebra. 
\lbl{Remark}
\end{Remark} 
The BRS variation of $\cT\zbz$ is obtained from Eq\rf{ct}
leading to :
 
\begin{eqnarray}
\sS \cT\zbz &=&\cC\zbz\prt\cT\zbz +2 \cT\zbz\prt\cC\zbz 
{-}
\cW\zbz \prt\cC^{(2)}\zbz\nn\\
&{-}&
\frac{2}{3}
\cC^{(2)}\zbz 
\prt\cW\zbz +\alpha\prt^3 \cC\zbz ,
\lbl{ct1}
\end{eqnarray}
in terms  of a spin $(3,0)$-conformal field $\cW$
whose BRS behaviour can be calculated from the nilpotency
condition applied to \rf{ct1},
\begin{eqnarray}
\sS \cW\zbz &=&\cC\zbz \prt \cW \zbz+ 3\prt\cC\zbz \cW\zbz
 + 
16
\cT\zbz
\prt\biggl(\cC^{(2)}\zbz\cT\zbz\biggr)\nn\\
&+&\alpha\biggl(\alpha\prt^5\cC^{(2)}\zbz
+
2
\cC^{(2)}\zbz\prt^3\cT\zbz
+10\cT\zbz\prt^3\cC^{(2)}\zbz\nn\\
&+&15\prt\cT\zbz\prt^2\cC^{(2)}\zbz
+
9
\prt^2\cT\zbz\prt\cC^{(2)}\zbz\biggr)
\lbl{scw}
\end{eqnarray}
The algebra here closes, since the nilpotency condition 
 holds whatever $\alpha$. 
In order to realize the role played by the parameter $\alpha$,
the stability under holomorphic changes of charts $z\lra w(z)$ of the
 algebra defined by the Eqs \rf{sc1} \rf{ct1} \rf{scw} is
now discussed. One has

\begin{eqnarray}
\cC^w = w'\cC^z,\qquad \cC^{ww} = (w')^2\cC^{zz},\qquad \prt_w =
\frac{1}{w'}\prt_z,
\end{eqnarray}
and the glueing rule of $\cT$ will be also worked out. Writing first
\rf{sc1} in the $w$ system of complex coordinates and then going back
to the $z$ one, we have
\begin{eqnarray}
\cC^w \prt_w \cC^w &=& w'\cC^z \frac{1}{w'}\prt_z (w'\cC^z ) = w'\cC^z
\prt_z \cC^z \nn\\
-\frac{16}{3} \cT_{ww} \cC^{ww} \prt_w \cC^{ww} &=& -\frac{16}{3}
(w')^3 \cT_{ww} \cC^{zz}\prt_z \cC^{zz} \nn
\end{eqnarray}
while the $\alpha$-term is more involved
\begin{eqnarray}
\alpha\biggl(\frac{1}{w'}\prt_z ((w')^2\cC^{zz}) \frac{1}{w'}\prt_z
[\frac{1}{w'}\prt_z  ((w')^2\cC^{zz})] -\frac{2}{3} (w')^2\cC^{zz}
\frac{1}{w'}\prt_z [\frac{1}{w'}\prt_z  (\frac{1}{w'}\prt_z
\{(w')^2\cC^{zz}\} )]\biggr) \nn\\
= \alpha w' \biggl( \prt_z \cC^{zz}\prt_z^2 \cC^{zz}
-\frac{2}{3}\cC^{zz}\prt_z^3 \cC^{zz}
-\frac{16}{3}\{w,z\}\cC^{zz}\prt_z \cC^{zz} \biggr)\nn
\end{eqnarray}
where $\{w,z\}$ denotes the Schwarzian derivative. Covariance requires
that
\begin{eqnarray}
\biggl((w')^3 \cT_{ww} + w' \alpha \{w,z\}\biggr)\cC^{zz}\prt_z \cC^{zz}
= w' \cT_{zz}\cC^{zz}\prt_z \cC^{zz}
\end{eqnarray}
so that
\begin{eqnarray}
(w')^2 \cT_{ww} + \alpha \{w,z\} = \cT_{zz}
\end{eqnarray}
showing that $\frac{1}{\alpha}\cT$ is a projective connection.
On the other hand if $\alpha=0$, $\cT$ is a tensor.
It is also easy to recover from Eqs \rf{ct1} \rf{scw}:
 \begin{eqnarray}
(w')^3 \cW_{www}  = \cW_{zzz}
\end{eqnarray}
showing that $\cW$ behaves as a true tensor of order three.
 
The solution can be found  if we notice that the parameter $\alpha $ can
be reabsorbed in the B.R.S. operator by rescaling:

 \begin{eqnarray}
\cC^{(2)}\zbz={\alpha}^{-\frac{1}{2}}\cC^{(2)}\zbz \nn\\ 
 \cT\zbz =\alpha\cT\zbz\nn\\  
 \cW\zbz={\alpha}^{\frac{3}{2}}\cW\zbz
\lbl{rescal}  
 \end{eqnarray}
 So if $\alpha\neq 0$ we can fix it to be equal to one without any trouble.
In the case $\alpha=1$ 
the B.R.S tranformations of $\cT\zbz$ and $\cW\zbz$ can be rewritten
 in terms of Bol derivatives \cite {Gieres}.

 \sect{$\cW_3$ algebra and coordinate tranformations:
 two different approaches for the same symmetry}

 The purpose of this Section is to discuss the algebra in Eq  \rf{sc1}
 \rf{ct1} \rf{scw}   
 found previously in terms of coordinate transformations.
 Indeed, we have  introduced the coordinates
  $Z^{(r)}\zbzt=Z^{(r)}_0\zbz+\theta Z^{(r)}_\theta\zbz, r=1,2$ in Eq \rf{zr}
  and  the  reparametrizations:

  \begin{eqnarray}
\zbz &\lra& \ZBZt\nn \\
\zbz &\lra& \ZBZduet
\lbl{diff2}
\end{eqnarray}
whose algebra under symplectomorphisms reads:
\begin{eqnarray}
\sS Z\zbzt &=&\cK^{(1)}\zbzt \prt Z\zbzt\nn\\
\sS Z^{(2)}\zbzt &=& \cK^{(2)}\zbzt \prt Z^{(2)}\zbzt
\end{eqnarray}
with ghosts:
\begin{eqnarray}
\cK^{(1)}\zbzt
 &=&\cC\zbz +\theta \cX^{(1)}\zbz
 \lbl{cK11} \\
\cK^{(2)}\zbzt
&=&\cC\zbz +\frac{{(\prt Z_0\zbz)}^2}{\prt Z_0^{(2)}\zbz}\cC^{(2)}\zbz\nn\\
&+&\theta \biggl[\cX^{(1)}\zbz  +
\underbrace{\frac{{(\prt Z_0\zbz)}^2}{\prt Z_0^{(2)}\zbz}
\cX^{(2)}\zbz}_{vanishing\quad in\quad \cW_3}\nn\\
&+& \biggl(2 \frac{\prt Z_\theta\zbz{\prt Z_0\zbz}}{\prt Z_0^{(2)}\zbz} 
- \frac{\prt Z^{(2)}_\theta\zbz{(\prt Z_0\zbz)}^2}{{(\prt
Z^{(2)}_0\zbz)}^2}\biggl)
\cC^{(2)}\zbz\biggr]
 \lbl{cK12}
\end{eqnarray}

The previous symmetry is broken at the $\zbz$ level, such that
the mappings:

\begin{eqnarray}
\zbz &\lra& \ZBZzero\nn \\
\lbl{diff10}
\zbz &\lra& \ZBZduezero
\lbl{diff20}
\end{eqnarray}
do not yield any reparametrization.

However a closed algebra can be written in terms of the BRS operation
acting on the previous coordinates: 

\begin{eqnarray}
\sS Z_0\zbz& =&\cC\zbz\prt Z_0\zbz +Z_\theta\zbz 
\lbl{sZ12a} \\
\sS Z_\theta\zbz
&=&\cC\zbz\prt Z_\theta\zbz  -\prt Z_0\zbz\cX^{(1)}\zbz
\lbl{sZ12b} \\ 
\sS Z^{(2)}_0\zbz &=&\cC\zbz \prt Z^{(2)}_0\zbz +\cC^{(2)}\zbz {(\prt
Z_0\zbz)}^2 +Z^{(2)}_\theta \zbz
\lbl{sZ12c}\\ 
\sS Z^{(2)}_\theta \zbz& =&\cC\zbz \prt Z^{(2)}_\theta\zbz -\cX^{(1)}\zbz \prt
Z^{(2)}_0\zbz \nn\\
&-& 2 \prt Z_0\zbz \prt Z_\theta\zbz \cC^{(2)}\zbz-\underbrace{\cX^{(2)}\zbz
{(\prt Z_0\zbz)}^2}_{vanishing\quad in\quad \cW_3} 
\lbl{sZ12d}
\end{eqnarray} 
where the $0$ label identifies good coordinate frames and the $\theta$
ones the breakings of the symmetry under reparametrizations. 
 
 So this algebra contains {\it anomalies} $\cX\zbz$ and
 $\cX^{(2)}\zbz$ which in the same formalism can be expressed  
 in terms of coordinates as: 
\begin{eqnarray}
\cX\zbz &: =& \cX^{(1)}\zbz \equiv \frac{\wdelta Z_\theta\zbz}{\prt Z_0\zbz}
\lbl{cX10} \\[2mm]
\cX^{(2)}\zbz&\equiv& 
\frac{1}{{(\prt Z_0\zbz)}^2}\biggl[\wdelta Z^{(2)}_\theta\zbz-\wdelta
Z_\theta\zbz \frac{\prt Z^{(2)}_0\zbz}{\prt Z_0\zbz}\nn\\ 
&-& \frac{2}{\prt Z_0\zbz}
 (\prt  Z_\theta\zbz)(\wdelta Z^{(2)}_0 -Z^{(2)}_\theta\zbz)\biggr]
\lbl{chi20}
\end{eqnarray}
where we have introduced the operator associated to $\delta_{[B.R.S.]}:=\sS$
\begin{eqnarray}
\wdelta\zbz\equiv\biggl[\delta_{[B.R.S.]}-\cC\zbz\prt\biggr] .
\end{eqnarray} 
These parameters are related to the breaking consistency conditions
$\wdelta Z_\theta\zbz$ and $\wdelta Z^{(2)}_\theta\zbz$:  since the
realization of the  $\cW_3$ algebra requires, as shown before, the
constraint  $\cX^{(2)}\zbz=0$,  we have to investigate in which
manner this  condition could affect   these
 transformations.  In the symplectic framework and  from Eqs
\rf{chi}\rf{cX2}, the previous condition will imply
 
\begin{eqnarray}
\wdelta Z^{(2)}_\theta\zbz=\wdelta Z_\theta\zbz \frac{\prt
Z^{(2)}_0\zbz}{\prt Z_0\zbz} + {2}\cC^{(2)}\zbz \prt Z_0\zbz \prt  Z_\theta\zbz
\lbl{w3cond}
\end{eqnarray}
when written in terms of coordinates.
 This formula encodes the geometrical content of the $\cW_3$ symmetry.

So this interplay between the breaking consitency conditions $\wdelta
 Z^{(2)}_\theta\zbz $, $\wdelta Z_\theta\zbz$ and coordinate
 transformations induces the symmetry   criterion  which  gives the origin
 to $\cW_3$ algebra through the condition $\cX^{(2)}\zbz=0$.
 Hence we can state:
\begin{statement} 
The condition  $\cX^{(2)}\zbz=0$ not only fixes the breaking terms of
 the diff algebra, 
 but also generates a mutual conspiracy between 
the breaking consistency conditions: 
the relationship between these violation mechanisms
generates the coordinate counterpart symmetry of the $\cW_3$
 algebra. 
\lbl{result} 
\end{statement}

\subsect{ Primary fields as explicit functions of coordinates}

This part of our paper wants to investigate an intriguing   aspect of our
 problem to which our approach gives a full meaning.
 
The conclusion reached in the previous statement \rf{result}
 transforms our algebra with the help of a differential complex for the
 infinitesimal transformations  written in terms of the coordinates 
 $Z_0\zbz$ and $Z^{(2)}_0\zbz$ in presence of breaking terms
 $Z_\theta\zbz$ and $Z^{(2)}_\theta\zbz$. 
From the physical (and even the mathematical) point of view a
 coordinate system represents a tool which is hard to do without.

Our approach embeds them in a $\zbz$ background whose presence gives a
meaning to our model;  nevertheless the {\it{explicit}} presence of
coordinates amounts to investigating the possibility   of writing
$Z_\theta\zbz$ and $Z^{(2)}_\theta\zbz$  as {\it{explicit}} expression
of both $Z_0\zbz$ and $\Zdzero$.

 We have seen before in the Ansatz \rf{Remark} that the Faddeev-Popov content
of the breaking terms can only be carried by the $\cC^{(2)}\zbz$ so we
argue as follows. Let us consider in a matrix-like notation the
following ansatz.
  
\begin{Ansatz}
\begin{eqnarray}
\left\{\begin{array}{c}
Z_\theta\zbz\\Z^{(2)}_\theta\zbz  
\end{array}\right\} 
=\sum_{r\geq 0} \prt^r \cC^{(2)}\zbz\left\{\begin{array}{c}
\cP^Z_{(2-r)}(\Zzero,\Zdzero)\\\cR^{Z^{(2)}}_{(2-r)}(\Zzero,\Zdzero) 
\end{array}\right\}
\lbl{Zthetadecomp}
\end{eqnarray}
\end{Ansatz} 
where $\cP^Z_{(2-r)}(\Zzero,\Zdzero)$ and
$\cR^{Z^{(2)}}_{(2-r)}(\Zzero,\Zdzero)$ are 
taken to be functions of both $Z_0\zbz$ and $\Zdzero$ and their
derivatives. One computes 

\begin{eqnarray}
\leqa{ \hspace{-7mm} 
\wdelta \left\{\begin{array}{c}
Z_\theta\zbz\\ Z^{(2)}_\theta\zbz  
\end{array}\right\}
= \sum_{r\geq 1}  \sum_{j=1}^r\left(\begin{array}{c} 
r\\[.5mm] j\end{array}\right)\prt^j\cC\zbz \prt^{r-j+1}\cC^{(2)}\zbz 
 \left\{\begin{array}{c}
 \cP^Z_{(2-r)}(\Zzero,\Zdzero)\\\cR^{Z^{(2)}}_{(2-r)}(\Zzero,\Zdzero) 
\end{array}\right\} }  \nn\\[2mm]
&& -2 \sum_{r\geq 0}\ \sum_{j=0}^r\left(\begin{array}{c} 
r\\[.5mm] j\end{array}\right)\prt^{j+1}\cC\zbz \prt^{r-j}\cC^{(2)}\zbz 
 \left\{\begin{array}{c}
 \cP^Z_{(2-r)}(\Zzero,\Zdzero)\\\cR^{Z^{(2)}}_{(2-r)}(\Zzero,\Zdzero) 
\end{array}\right\} \nn\\[2mm]
& & -\ \sum_{r\geq 0} \left( \prt^r \cC^{(2)}\zbz
 \sum_{n\geq 0} \biggl[ \frac{ \prt
 }{ \prt ({\prt^n} Z_0\zbz)} \left\{\begin{array}{c}
 \cP^Z_{(2-r)}(\Zzero,\Zdzero)\\\cR^{Z^{(2)}}_{(2-r)}(\Zzero,\Zdzero) 
\end{array}\right\}  \right. \nn\\[2mm]
 && \qquad {\prt^n} \biggl( \cC\zbz \prt Z_0\zbz + 
 \sum_{s\geq 0} \prt^s \cC^{(2)}\zbz \left\{\begin{array}{c}
 \cP^Z_{(2-s)}(\Zzero,\Zdzero)\\ \cP^Z_{(2-s)}(\Zzero,\Zdzero)
\end{array}\right\} \biggr) \nn\\[2mm]
 && +\ \frac{ \prt
 }{\prt( {\prt^n} Z^{(2)}_0\zbz)}\left\{\begin{array}{c}
 \cP^Z_{(2-r)}(\Zzero,\Zdzero)\\\cR^{Z^{(2)}}_{(2-r)}(\Zzero,\Zdzero) 
\end{array}\right\}
{\prt^n} \biggl( \cC\zbz \prt Z^{(2)}_0\zbz\nn\\[2mm]
 &&\qquad +\ \cC^{(2)}\zbz{(\prt Z_{(0)}\zbz)}^2
 +\sum_{s\geq 0} \prt^s \cC^{(2)}\zbz \left\{\begin{array}{c}
\cR^{Z^{(2)}}_{(2-s)}(\Zzero,\Zdzero) \\ \cR^{Z^{(2)}}_{(2-s)}(\Zzero,\Zdzero) 
\end{array}\right\}
\biggr) \biggr]  \nn\\[2mm]
&& \left. -\ \cC\zbz \prt \left\{\begin{array}{c}
 \cP^Z_{(2-r)}(\Zzero,\Zdzero)\\\cR^{Z^{(2)}}_{(2-r)}(\Zzero,\Zdzero) 
\end{array}\right\} \right) 
\lbl{wdelta}
\end{eqnarray}
The previous $\wdelta$ variations have Faddeev-Popov charge equal to
 two and contain both\\  $\prt^s\cC\zbz\prt^r\cC^{(2)}\zbz$ and
 $\prt^n\cC^{(2)}\zbz\prt^m\cC^{(2)}\zbz$ ghost monomials.  In order
 to implement Eqs\rf {sZ12b} and \rf{sZ12d}, recall that in the solution
 \rf{X0} for $\cX\zbz$,   all the mixed terms
 $\prt^s\cC\zbz\prt^r\cC^{(2)}\zbz$ have to cancel out.  This gives
 rise to the following set of constraints on both the unknowns $\cP$
 and $\cR$. It will be useful for the sequel to proceed as follows;
 we have two distinct systems of conditions. First 
 the vanishing of the coefficient of the monomials
$\cC^{(2)}\prt^{r+1}\cC$, $r\geq 0$, yields
\begin{eqnarray}
\left\{\begin{array}{c}
\cP^Z_{(2-r)}(\Zzero,\Zdzero)\\\cR^{Z^{(2)}}_{(2-r)}(\Zzero,\Zdzero) 
\end{array}\right\}
=  \frac{1}{2 }\,  \cV^{(r)}
\left\{\begin{array}{c}
\cP^Z_{(2)}(\Zzero,\Zdzero)\\\cR^{Z^{(2)}}_{(2)}(\Zzero,\Zdzero) 
\end{array}\right\}, \quad r\geq 0,
 \lbl{vanish1} 
\end{eqnarray}
and the one for the monomials 
$\prt^r\cC^{(2)}\prt^{s+1}\cC$, $r \geq 1$, $s\geq 0$
 gives rise to
\begin{eqnarray}
\leqa{ \left\{\begin{array}{c}
\cP^Z_{(2-(r+s))}(\Zzero,\Zdzero) \\
\cR^{Z^{(2)}}_{(2-(r+s))}(\Zzero,\Zdzero)
\end{array}\right\} = } \nn\\
&=&\frac{r!(s+1)!}{(s+r)!(2(s+1)-r)}\; \cV^{(s)}\left\{\begin{array}{c}
\cP^Z_{(2-r)}(\Zzero,\Zdzero)\\\cR^{Z^{(2)}}_{(2-r)}(\Zzero,\Zdzero)
\end{array}\right\}
\lbl{vanish2}\\
&&\mbox{for } r \geq 1,\ s\geq 0, \mbox{ and } r\neq 2(s+1), \nn
\end{eqnarray}
where we have set for $r\geq 0$:
\begin{eqnarray}
\cV^{(r)}  
     = \hspace{ -3mm}
\sum_{n\geq (r+1)} \left(\begin{array}{c} 
n\\r+1\end{array}\right) \biggl[  \prt^{n-r} Z_0\zbz \frac{ \prt
}{ \prt(\prt^n Z_0\zbz)}
+  \prt^{n-r}Z^{(2)}_0\zbz 
 \frac{ \prt
}{ \prt(\prt^n Z^{(2)}_0\zbz)}
  \biggr]
  \lbl{cV} 
\end{eqnarray}
It is worthwhile to remark that $\cV^{(0)}$ is the counting
operator with respect to the little indices $z$.
Note however that the conditions \rf{vanish1} is contained in
\rf{vanish2} by allowing for $r=0$. This decomposition of the
constraints leads advantageously to the following
 
\begin{statement}
Owing to Eq.\rf{vanish2}, for $r, s\geq 0$, the  functions
$\cP^Z_{(2-(r+s))}(\Zzero,\Zdzero)$ 
and $\cR^{Z^{(2)}}_{(2-(r+s)))}(\Zzero,\Zdzero)$ can be completely fixed 
in terms of $\cP^Z_{(2-r)}(\Zzero,\Zdzero)$ and 
$\cR^{Z^{(2)}}_{(2-r))}(\Zzero,\Zdzero)$, respectively, in an independent
way. 
 
The operator $\cV^{(r)}$ defined in Eq\rf{cV}  decreases   the order of all
 derivatives by $r$. Thus the construction of $\cP^Z_{(2-r)}(\Zzero,\Zdzero)$ 
in terms of  $\cP^Z_{(2)}(\Zzero,\Zdzero)$ means that,
 if $\tld{m}$ is the highest order of 
  derivative terms of $Z\zbz$ or $Z^{(2)}\zbz$ contained in
 $\cP^Z_{(2)}(\Zzero,\Zdzero)$, then the functions  
  $\cP^Z_{(2-\tld{m}-r)}(\Zzero,\Zdzero)$, $r>1$ will be zero.
   
   In particular, any $\cV^{(r)}$ for $r \geq 1$ gives no action on 
$\cP^Z_{(2)}(\Zzero,\Zdzero)$  involving zero and first order derivative terms.
  The same argument holds for  $\cR^{Z^{(2-r)}}_{(2))}(\Zzero,\Zdzero)$ in
  relation to $\cR^{Z^{(2)}}_{(2))}(\Zzero,\Zdzero)$. 
\end{statement} 

An important remark is in order.
By acting with $\cV^{(r)}$ on the lower
 states $\cP^Z_{(2)}(\Zzero,\Zdzero)$ and $\cR^{Z(2)}_{(2)}(\Zzero,\Zdzero)$
 all the upper states can be constructed. But due to \rf{vanish2}
the states $\cP^Z_{(2-(r+s))}(\Zzero,\Zdzero)$ and
 $\cR^{Z(2)}_{(2-(r+s))}(\Zzero,\Zdzero)$ are obtained from
$\cP^Z_{(2-r)}(\Zzero,\Zdzero)$ and
$\cR^{Z(2)}_{(2-r)}(\Zzero,\Zdzero)$ 
acting with $\cV^{(s)}$ only if $r\neq 2(s+1)$. The latter
 limits  the cancellation mechanism of the $\prt^r\cC^{(2)}\prt^{s+1}\cC$
  monomials in Eq.\rf{wdelta} only to selected states (since the sum
$(r+s)$ can be reached in a non unique way without
spoiling the inequality $r  \neq 2(s+1)$) in the  decomposition 
\rf{Zthetadecomp}.

The previous cancellation mechanisms could hide an {\em a priori}
inconsistency:   this is not the case. 
In fact  the  local operators   $\cV^{(r)}$ defined in \rf{cV} verify
the algebra 

\begin{eqnarray}
\biggl[\cV^{(r)},\cV^{(s)}\biggr]=(r-s)
\frac{(r+s+1)!}{(r+1)!(s+1)!} 
\,\cV^{(r+s)}  ,
\lbl{commcV}
\end{eqnarray}
so that Eqs\rf{vanish1} and \rf{vanish2} can be  related.
Indeed, combining both Eqs\rf{vanish1} and \rf{commcV} one obtains 

\begin{eqnarray}
\leqa{ \hskip -1cm \left\{\begin{array}{c}
\cP^Z_{(2-(r+s))}(\Zzero,\Zdzero) \\
\cR^{Z^{(2)}}_{(2-(r+s))}(\Zzero,\Zdzero)
\end{array}\right\}\ =\
 \frac{\cV^{(r+s)}}{2}
\left\{\begin{array}{c}
\cP^Z_{(2)}(\Zzero,\Zdzero) \\
\cR^{Z^{(2)}}_{(2)}(\Zzero,\Zdzero)
\end{array}\right\} }  \nn\\
&=&\frac{(s+1)!(r+1)!}{(s+r+1)!(s-r)}\Biggl( \cV^{(s)}\left\{\begin{array}{c}
\cP^Z_{(2-r)}(\Zzero,\Zdzero)\\\cR^{Z^{(2)}}_{(2-r)}(\Zzero,\Zdzero)
\end{array}\right\} \nn\\
&& \hskip 5.cm -\; \cV^{(r)}\left\{\begin{array}{c}
\cP^Z_{(2-s)}(\Zzero,\Zdzero)\\\cR^{Z^{(2)}}_{(2-s)}(\Zzero,\Zdzero)
\end{array}\right\}\Biggr)\nn
\end{eqnarray}
 Using Eq.\rf{vanish2} we get an identity which holds for $r \neq 2(s+1)$ and
$s \neq 2(r+1)$. The latter formula shows moreover that
$\cP^Z_{(2-(r+s))}$ and/or $\cR^{Z^{(2)}}_{(2-(r+s))}$ can be obtained
without any restriction. Hence \rf{wdelta} reduces to
\begin{eqnarray}
\wdelta \left\{\!\!\!\!\begin{array}{c}
Z_\theta\zbz\\Z^{(2)}_\theta\zbz  
\end{array}\!\!\!\!\right\} 
= \sum_{r,s\geq 0} \prt^s\cC^{(2)}\zbz \prt^r\cC^{(2)}\zbz 
\cL_{(2-s)}\zbz \,\frac{\cV^{(r)}}{2}
 \left\{\!\!\!\!\begin{array}{c}
\cP^Z_{(2)}(\Zzero,\Zdzero)\\\cR^{Z^{(2)}}_{(2)}(\Zzero,\Zdzero) 
\end{array}\!\!\!\!\right\}
\lbl{rigidcond}
\end{eqnarray}
where the differential operator $\cL_{(2-s)}$ is defined by
 \begin{eqnarray}
\cL_{(2-s)}\zbz &:=&
\sum_{\footnotesize \begin{array}{c} k,l \geq 0 \\ k+l =s \end{array} }  
\sum_{n\geq k} \left( \begin{array}{c} n\\ k \end{array} \right) 
\left( \biggl( \prt^{n-k} \frac{\cV^{(l)}}{2}\cP_{(2)}(\Zzero,\Zdzero)\biggr)
\frac{\prt}{ \prt(\prt^n Z_0\zbz)} \right. \nn\\ 
&&+\ \left. \biggl( \prt^{n-k}\frac{\cV^{(l)}}{2}
 \cR^{Z^{(2)}}_{(2)}(\Zzero\Zdzero) 
\biggr) \right)  \frac{\prt}{ \prt(\prt^n Z^{(2)}_0\zbz)}\nn\\
&&+\ \sum_{n\geq s} \left(\begin{array}{c} n\\ s\end{array} \right) 
\biggl( \prt^{n-s} (\prt Z_0)^2\zbz\biggr) \frac{\prt}{\prt(\prt^n
 Z^{(2)}_0\zbz)} 
 \lbl{cL}
 \end{eqnarray} 
 The $\cL_{(2-s)}\zbz$ operator looks like a coordinate tranformation operator  wich maps $\Zzero$ and its derivatives 
 into  particular derivatives of  $\cP_{(2-l)}(\Zzero,\Zdzero)$  and similarly for  $\Zdzero$  with respect 
  $\cR^{Z^{(2)}}_{(2-l)}(\Zzero,\Zdzero) $ plus an inhomogeneous 
term depending on $\Zzero$. 
 
But $\wdelta Z_\theta\zbz $  has a well defined Faddeev-Popov ghost
content due to the condition \rf{X0}. 
Only the $\cC^{(2)}\zbz\prt \cC^{(2)}\zbz $ ,$\cC^{(2)}\zbz\prt^3
\cC^{(2)}\zbz$, and $\prt\cC^{(2)}\zbz\prt^2 \cC^{(2)}\zbz $ ghost monomials
are present in the expansion \rf{rigidcond}: 
the  first term  will provide  a $\Zzero$, $\Zdzero$ for the $\cT$ function; while the  
other ones give the $\alpha$ term in Eq \rf{X0}.
All the other  terms must be zero.

These conditions provide a  system of equations  for $\cP$ and $\cR$.

A capital role  is played by the equations for $\alpha$ terms,  
since they provide a coordinate representation for the constant
$\alpha$ which has to be true for each $\Zzero$, $\Zdzero$ change of
charts. This covariance requirement, combined with  
the algebraic stucture of the two local operators $\cV^{(s)}$ and
 $\cL_{(2-s)}$   show that  only the value  $\alpha =0$ is consistent for this approach.
 
 \begin{eqnarray}
\cL_{(2-s)} \cP_{(2-r)}(\Zzero,\Zdzero) = \cL_{(2-r)} \cP_{(2-s)}(\Zzero,\Zdzero) \nn\\
(r+s)\neq 1\nn\\
\lbl{lp}
\end{eqnarray} 

This  equation can be treated  as an integrability  condition  and contains  both a linear and a bilinear part 
in the  $\cP_{(2-r)}(\Zzero,\Zdzero)$ and the $\cR^{Z^{(2)}}_{(2-r)}(\Zzero,\Zdzero) $  functions.

Filtering with the counting  operators  of the previous functions    we select the linear part of Eq
\rf{lp}.
So we have an infinite set of conditions on  $\cP_{(2-r)}(\Zzero,\Zdzero)$ 

For $r=0$ , $s>1$ we get:
 \begin{eqnarray}
\cL^{lin}_{(2-s)} \cP_{(2)}(\Zzero,\Zdzero)=\cL^{lin}_{(2)} \cP_{(2-s)}(\Zzero,\Zdzero) \nn\\
\lbl{lp0}
\end{eqnarray} 

where:
\begin{eqnarray}
\cL^{lin}_{(2-s)}=\sum_{n\geq s} \left(\begin{array}{c} n\\ s\end{array} \right) 
\biggl( \prt^{n-s} (\prt Z_0)^2\zbz\biggr) \frac{\prt}{\prt(\prt^n
 Z^{(2)}_0\zbz)} 
\end{eqnarray}

This condition must be compatible with the $\cV^{(r)}$ algebra stated in the equation \rf{vanish1} for $\cP_{(2)}(\Zzero,\Zdzero)$.

 The form of the operator $  \cL^{lin}_{(2-s)}$  forbids this possibility; the only way to escape is to 
 prevent the dependence of these function 
on the derivatives of $\Zdzero$ with order greater or equal than two.  Going on with the same  filtration mechanism 
we forbid the dependence on the 
derivatives of $\Zzero $  of the same order.

  This shows that $Z_\theta\zbz $  depends only on the derivatives of $\Zzero$ and $\Zdzero$ of order less than two: 
  now the same argument can be repeteated  for $\wdelta Z^{(2)}_\theta\zbz  $ in the same Faddeev-Popov  and derivatives sectors as before;  
 so we obtain an analogous equation as Eq \rf{lp} for the $\cR^{Z^{(2)}}_{(2-r)}(\Zzero,\Zdzero) $ functions  obtaining the same results as before 
 for the $\cR^{Z^{(2)}}_{(2)}(\Zzero,\Zdzero) $ functions.
 
 So we get:

\begin{statement} 
 The only non zero 
functions in Eq \rf{Zthetadecomp} will be
$\cP^Z_{(2)}(\Zzero,\Zdzero)$, and $\cR^{Z^{(2)}}_{(2)}(\Zzero,\Zdzero)$.

The other functions will be zero if, as stated by Eq \rf{vanish1}, 
the previous functions depend only on $\Zzero$ and $\Zdzero$
 and their first order derivatives. 
\lbl{statement}
\end{statement}

This means  that $\cP^Z_{(2)}(\Zzero,\Zdzero)$,
   $\cR^{Z^{(2)}}_{(2)}(\Zdzero,\Zdzero)$ and $\cT\zbz$ are well-defined
    tensors under holomorphic change of charts.

Assuming  the general  expressions:
\begin{eqnarray}
&&\cP^Z_{(2)}(\Zzero,\Zdzero)
=\sum_i A_i 
\prod_j{\biggl( \prt^{m^i_j}\Zzero\biggr)}^{{\beta}^i_j}
\biggl( \prt^{n^i_j}\Zdzero\biggr)}^{{\sigma}^i_j}
\bigg|_{\tiny{\left\{
\begin{array}{c}
{\sum_j \beta^i_j+2\sigma^i_j=1}\\
{\sum_j\beta^i_j m^i_j+\sigma^i_j n^i_j =2}\\
{\forall i,j} 
\end{array}\right\}} \nn\\
&& \hskip 5cm
f_i\biggl(
\frac{Z^{(2)}_0\zbz}{{(Z_0\zbz)}^2},\frac{\prt Z^{(2)}_0\zbz}{\prt
{(Z_0\zbz)}^2} 
\biggr)
 \lbl{pform}
\end{eqnarray} 
\begin{eqnarray}
&&\cR^{Z^{(2)}}_{(2)}(\Zzero,\Zdzero)
=\sum_i B_i 
\prod_j{\biggl( \prt^{l^i_j}\Zzero\biggr)}^{{\rho}^i_j}
\biggl( \prt^{h^i_j}\Zdzero\biggr)}^{{\psi}^i_j}
\arrowvert
_{\tiny{\left\{
\begin{array}{c}
{\sum_j \rho^i_j+2\psi^i_j=2}\\
{\sum_j\rho^i_j l^i_j+\psi^i_j h^i_j =2}\\
{\forall i,j} 
\end{array}\right\}} \nn\\
&&\hskip 5cm
g_i\biggl(
\frac{Z^{(2)}_0\zbz}{{(Z_0\zbz)}^2},\frac{\prt Z^{(2)}_0\zbz}{\prt
{(Z_0\zbz)}^2}
\biggr)  
\lbl{rform}
\end{eqnarray} 
where  $f_i,g_i $ are arbitrary scalar functions.

with the constraints imposed  in Eq \rf{pform},\rf{rform}  by the Statement \rf{statement}:
\begin{eqnarray}
m^i_j,   n^i_j,  l^i_j,      h^i_j =0,1
\end{eqnarray}
At this stage the role of Eq \rf{w3cond} appears as the condition which
links together the functions
$\cP^Z_{(2-r)}(\Zzero,\Zdzero)$ and $\cR^{Z^{(2)}}_{(2-r)}(\Zzero,\Zdzero)$
since:
\begin{eqnarray}
 && \prt Z_0\zbz
 \cL_{(1)}\zbz
 \cR^{Z^{(2)}}_{(2)}(\Zzero,\Zdzero)\nn\\
&=&{\prt Z^{(2)}_0\zbz}
\cL_{(1)}\zbz
 \cP^Z_{(2)}(\Zzero,\Zdzero)  \nn\\
&+& {2}{(\prt Z_0\zbz)}^2
 (\cP^Z_{(2)}(\Zzero,\Zdzero))\nn\\ 
\lbl{w3cond1}
\end{eqnarray}
 we get 
\begin{eqnarray}
 \cP^Z_{(2)}(\Zzero,\Zdzero)&=&A \frac{{(\prt Z_0\zbz)}^3}{\prt Z^{(2)}_0\zbz}\\
  \cR^{Z^{(2)}}_{(2)}(\Zzero,\Zdzero)&=&B {(\prt Z_0\zbz)}^2\nn\\
  B=A-1
  \lbl{solution1} 
  \end{eqnarray} 
so:
 \begin{eqnarray}
\sS Z_0\zbz& =&\biggl(\cC\zbz+A\cC^{(2)}\zbz\frac{{(\prt Z_0\zbz)}^2}{\prt
Z^{(2)}_0\zbz}\biggr) \prt Z_0\zbz\nn \\
\lbl{sZ121a}\nn\\
\sS Z^{(2)}_0\zbz &=&\biggl(\cC\zbz+A\cC^{(2)}\zbz\frac{{(\prt Z_0\zbz)}^2}{\prt
Z^{(2)}_0\zbz}\biggr)  \prt Z^{(2)}_0 \zbz\nn\\
\lbl{sZ121c}
\end{eqnarray}

so that for a rescaling $ \cC^{(2)}\zbz\lra\frac{1}{ A}\cC^{(2)}\zbz$ 
we find a perfect covariance of $Z_0\zbz$  $ Z^{(2)}_0\zbz$ 
under the ghost $\cK_2\zbz =\biggl(\cC\zbz+\cC^{(2)}\zbz\frac{{(\prt
Z_0\zbz)}^2}{\prt Z^{(2)}_0\zbz}\biggr)  $
 
Finally we find:
\begin{eqnarray}
\cT\zbz &=&\frac{3}{8}{\biggl( \frac{{(\prt Z_0\zbz)}^2}{\prt Z^{(2)}_0\zbz
}\biggr)}^2
\lbl{alpha3}\\
\cW\zbz &=&-\frac{3}{4}{\biggl( \frac{{(\prt Z_0\zbz)}^2}{\prt Z^{(2)}_0\zbz
}\biggr)}^3
\lbl{alpha0}
\end{eqnarray}

Indeed it is easy to verify that the Eq \rf{w3cond1} is fully equivalent to say
that the $\theta$ part of Eq\rf{cK12} becomes zero.

The explicit dependence on the coordinates reconstructs the diffeomorphism
symmetry.

In the case of the algebra Eq\rf{grimm} if we put:
\begin{eqnarray}
 Z_\theta&=&A \cC^{(n)}\zbz\frac{{(\prt Z_0\zbz)}^{(n+1)}}{\prt Z^{(n)}_0\zbz}\\
  Z^{(n)}_\theta\zbz&=&B {(\prt Z_0\zbz)}^{(n)}, \mbox{ with } B=A-1
  \lbl{solutiongrimm} 
  \end{eqnarray} 
we find the covariance
\begin{eqnarray}
\sS Z_0\zbz& =&\biggl(\cC\zbz+\cC^{(n)}\zbz\frac{{(\prt Z_0\zbz)}^{(n+1}}{\prt
Z^{(n)}_0\zbz}\biggr) \prt Z_0\zbz \\
\lbl{sZ1n1a}
\sS Z^{(n)}_0\zbz &=&\biggl(\cC\zbz+\cC^{(n)}\zbz\frac{{(\prt
Z_0\zbz)}^{(n+1)}}{\prt Z^{(n)}_0\zbz}\biggr)  \prt Z^{(n)}_0 \zbz
\lbl{sZ1n1c}
\end{eqnarray}
under the ghost field
\begin{eqnarray}
\cK^{(n)}\zbz =\biggl(\cC\zbz+\cC^{(n)}\zbz\frac{{(\prt Z_0\zbz)}^{(n+1)}}{\prt
Z^{(n)}_0\zbz}\biggr)  
\end{eqnarray}

\subsect{Advantages and drawbacks of the different approaches }

In this, we will point out the worths and the faults of choosing the
primary fields as explicit functions of the coordinates or not. 

First of all the coordinate description honors its scoreboard with
the explicit residual reparametrization  symmetry $\zbz \lra \ZBZdue$. 
The price to pay for it is to lose in the way the $\cT\zbz$ fiels as a
projective connection. 
These geometrical objects are welcome in conformal  ``covariant''
theories, when the holomorphic  derivative operators  are  
to be defined in a covariant way.
 
It is a well-known  problem for a right  definition of the
 diffeomorphism anomaly  \cite{Gieres}. 
 Its solution is quite involved and its origin is not a clearly  identified.

 We want to show that the anomaly of our algebra is, in the {\it
 explicit coordinate} approach  ($\alpha=0$)  not naively well
 defined, so that complicated surgery methods  have to be applied, while
 if $\cT\zbz$ can be considered as a projective connection($\alpha\neq0$)
 , the anomaly becomes well defined under change of charts.

We define the anomaly in terms of Gel'fand-Fuchs cocycles as in
reference \cite{BaLa93}.

The holomorphic cocycles $\Delta^{(\natural,n)}\zbz$ of Faddeev-Popov
charge equal to $n$ is defined as:  
 
\begin{eqnarray}
\sS\Delta^{(\natural,n)}\zbz=0
\lbl{holoc}
\end{eqnarray}

So if we  decompose:
\begin{eqnarray}
\Delta^{(\natural, n)}\zbz=\cC\zbz \Delta^{(n-1)}_z \zbz
+\widehat{\Delta^n_0}\zbz
\end{eqnarray}
Eq. \rf{holoc}  implies:
\begin{eqnarray}
\wdelta\zbz \Delta^{(n-1)}_z \zbz -\prt\cC\zbz\Delta^{(n-1)}_z \zbz  
-\prt\widehat{\Delta^n_0}\zbz=0\nn\\
\wdelta\zbz\widehat{\Delta^n_0}\zbz+\cX\zbz \Delta^{(n-1)}_z \zbz =0
\end{eqnarray}
The Feigin-Fuchs cocycle is the Faddeev-Popov 
charge three well defined element 
of the cohomology space.

Somewhat elaborated calculations will give:

\begin{eqnarray}
{\Delta}^3_0\zbz &=&(\sS \cC\zbz)\prt^2\cC\zbz +\prt(\cC\zbz\prt \cC\zbz
 -\sS \cC\zbz)\prt \cC\zbz\nn\\[2mm] 
&-&\frac{2}{3}\cW\zbz\Ch\zbz\prt\Ch\zbz\prt^2\Ch\zbz\nn\\[2mm]
&+& \sm{1}{3}(\alpha(\prt^2\Ch\zbz\prt^3\Ch\zbz - \prt\Ch\zbz\prt^4\Ch\zbz
)\nn\\[2mm]
& +&
2\prt^2\cT\zbz\Ch\zbz\prt\Ch\zbz 
-10\cT\zbz\prt\Ch\zbz\prt^2\Ch\zbz\nn \\[2mm] 
& - &2\prt\cT\zbz\Ch\zbz\prt^2\Ch\zbz)\cC\zbz
\lbl{ff}
\end{eqnarray}

Going at $\alpha=0$ and substituting the explicit expression of $\cT\zbz$ and
$\cW\zbz$ as in Eqs \rf{alpha3}\rf{alpha0}
 we recover the {\it{explicit coordinate}} description, so  Eq\rf{ff} can be
written in term of the 
$\cK^{(2)}\zbz$ as:

\begin{eqnarray}
{\Delta}^3_0\zbz &=&\cK^{(2)}\zbz \prt\cK^{(2)}\zbz\prt^2\cK^{(2)}\zbz 
\end{eqnarray}

The anomaly can be calculated along the lines found in Ref \cite{Ba88,BaLa93}:
 
\begin{eqnarray}
&&{\Delta}^{1}_{2}\zbz dz\wedge d\bz= \frac{\prt}{\prt c\zbz}\frac{\prt}{\prt
\bc\zbz}
\Delta^3_0\zbz dz\wedge d\bz\nn\\
&=&\prt\mu\zbz \prt^2\cC\zbz -\prt\cC\zbz\prt^2\mu\zbz\nn\\
 &+&\frac{1}{3}\biggl(\alpha(\prt^3\mu^{(2)}\zbz\prt^3 \cC^{(2)}\zbz
 -\prt^2\cC^{(2)}\zbz\prt^3\mu^{(2)}\zbz
 \nn\\
 &-&
 \prt\mu^{(2)}\zbz
\prt^4\cC^{(2)}\zbz+\prt\cC^{(2)}\zbz\prt^4\mu^{(2)}\zbz)\nn\\
 &+&2\prt^2\cT\zbz(\mu^{(2)}\zbz\prt\cC^{(2)}\zbz
-\cC^{(2)}\prt\mu{(2)}\zbz\biggr)\nn\\
&-&10\cT\zbz(\prt\mu^{(2)}\zbz\prt^2\cC^{(2)}\zbz-\prt\cC^{(2)}\zbz\prt^2\mu^{(2)}\zbz)
 \nn\\
&-&2\prt\cT\zbz(\mu^{(2)}\zbz\prt^2\cC^{(2)}\zbz-\cC^{(2)}\zbz\prt^2\mu^{(2)}\zbz
\lbl{anomaly}
 \end{eqnarray} 
where the (extended) Beltrami multipliers are
\cite{geom}\cite{Sorella}\cite{Grimm} \cite{BaLa00}:
\begin{eqnarray}
\mu\zbz=\frac{\prt \cC\zbz}{\prt \bc\zbz}\quad
\mu^{(2)}\zbz=\frac{\prt \cC^{(2)}\zbz}{\prt \bc\zbz}
\end{eqnarray} 
for $\alpha=0$ we get the usual anomaly 
\begin{eqnarray}
{\Delta}^{1}_{2}\zbz dz\wedge d\bz=\cK^{(2)}\zbz\prt^3\biggl(\mu\zbz
+\frac{{(\prt Z_0\zbz)}^2}{\prt Z^{(2)}_0\zbz} \mu^{(2)}\zbz\biggr)
\end{eqnarray} 
since ${\dps \mu\zbz +\frac{{(\prt Z_0\zbz)}^2}{\prt Z^{(2)}_0\zbz}
\mu^{(2)}\zbz}$ 
is the Beltrami multiplier of $Z^{(2)}_0\zbz$ with respect the $\zbz$
background. 

This object is not well defined under holomorphic changes of charts
\cite{becchi} 
\cite{Gieres}, and an intricate game, necessary to introduce a projective
connection
 (useful for geometrical purposes but unessential for the dynamical ones) is
needed.

On the other hand for $\alpha\neq0$, (so we can fix $\alpha=1$ due to 
  the rescaling property Eq \rf{rescal}) a projective connection $\cT\zbz$ is at
our disposal.
 So,   adding total derivatives and cocycles to Eq\rf{anomaly}(unessential from
the cohomological but essential from the geometrical
point of view) we can rewrite:
 \begin{eqnarray}
&& \Delta^{1}_{2}\zbz dz\wedge d\bz=\biggl\{\biggl(\cC\zbz
L_3\mu^z_\bz\zbz-\mu^z_\bz\zbz L^3\cC\zbz\biggr)\nn\\
& -&\frac{1}{3}\biggl(\cC^{(2)}\zbz L_5\mu^{(2)}_\bz\zbz -\mu^{(2)}_\bz\zbz
L^5\cC^{(2)}\zbz \biggr)\nn\\
 &-&8\biggl(\cC\zbz \mu^{(2)}\zbz -\mu^z_\bz\cC^{(2)}\zbz\biggr)\cW\zbz\nn\\
 &-&24\cW\zbz\biggl(\cC\zbz\prt\mu^{(2)}_\bz\zbz-\mu^{(2)}_\bz\zbz\prt\cC\zbz
\nn\\
 &+&\cC^{(2)}\zbz \prt\mu^z_\bz\zbz
-\mu^z_\bz\zbz\prt\cC^{(2)}\zbz\biggr)\biggr\}dz\wedge d\bz
\lbl{anomalygood}
 \end{eqnarray}

where the Bol derivatives are recalled to be:
\begin{eqnarray}
L_3& =& \prt^3 +2\cT\zbz \prt +(\prt\cT\zbz)\nn \\
L_5 &=&\prt^5 +10 \cT\zbz \prt^3 +15(\prt\cT\zbz)\prt^2 \nn\\
 &+&\biggl[ 9(\prt\cT\zbz) +16 \cT^2\zbz \biggr]\prt +2\biggl[(\prt^3\cT\zbz
+8\cT\zbz (\prt\cT\zbz)\biggr]
\end{eqnarray}

We see that the anomaly is well defined under holomorphic change of
charts, due to the fact 
that $\cT\zbz$ is a projective connection.

\sect{Conclusions}

We have shown in this paper the geometrical origin and  the strengths
and weaknesses of $\cW_3$  and the other algebras related
to the same construction.

The symplectic approach to this problem fits
in a coordinate scenario  the O.P.E   method   from which,
historically speaking, the $\cW$ algebras were derived.  There are more
questions that the reader can ask. 
 What deeper insights
will lock the symplectic geometry?
Is there a further relation between O.P.E.  and the symmetry
constraints?  The answers could reveal  the connections  between
Physics and Geometry, but unfortunately they are not at our hand.

{\bf{Aknowledgements}} We would like to thank Alberto Blasi for
several discussions and patience.


\end{document}